# Topological Supercavity Resonances In The Finite System

**Lujun Huang**[1#*], **Bin Jia**[2#], **Yan Kei Chiang**[1], **Sibo Huang**[2], **Chen Shen**[3], **Fu Deng**[1], **Tianzhi Yang**[4], **David A Powell**[1], **Yong Li**[2*], and **Andrey E Miroshnichenko**[1*]

[1] *School of Engineering and Information Technology, University of New South Wales, Canberra, Northcott Drive, ACT, 2600, Australia*

[2] *Institute of Acoustics, Tongji University, Shanghai, 200092, People's Republic of China*

[3] *Department of Mechanical Engineering, Rowan University, Glassboro, NJ, 08028, USA*

[4] *School of Mechanical Engineering and Automation, Northeastern University, Shenyang, 110819 China*

\# These authors contributed equally to this work

[*] *lujun.huang@unsw.edu.au, yongli@tongji.edu.cn, andrey.miroshnichenko@unsw.edu.au*

Acoustic resonant cavities play a vital role in modern acoustical systems. They have led to many essential applications for noise control, biomedical ultrasonics, and underwater communications. The ultrahigh quality-factor resonances are highly desired for some applications like high-resolution acoustic sensors and acoustic lasers. Here, we theoretically propose and experimentally demonstrate a new class of supercavity resonances in a coupled acoustic resonators system, arising from the merged bound states in the continuum (BICs) in geometry space. We demonstrate their topological origin by explicitly calculating their topological charges before and after BIC merging, accompanied by charges annihilation. Comparing with other types of BICs, they are robust to the perturbation brought by fabrication imperfection. Moreover, we found that such supercavity modes can be linked with the Friedrich-Wintgen BICs supported by an entire rectangular (cuboid) resonator sandwiched between two rectangular (or circular) waveguides, and thus more supercavity modes are constructed. Then, we fabricate these coupled resonators and experimentally confirm such a unique phenomenon: moving, merging, and vanishing of BICs by measuring their reflection spectra, which show good agreement with the numerical simulation and theoretical prediction of mode evolution. Finally, given the similar wave nature of acoustic and electromagnetic waves, such merged BICs also can be constructed in a coupled photonic resonator system. Our results may find exciting applications in acoustic and photonics, such as enhanced acoustic emission, filtering, and sensing.

# 1. Introduction

Acoustic resonators constitute the fundamental element of many modern acoustic devices. They can be used as unit cells of acoustic metasurface that allow for arbitrary manipulation of acoustic wavefront[1,2], leading to practical applications including acoustic absorbers[3] and invisibility cloaking[4]. The most typical example is the Helmholtz resonator that has been widely used in the sound absorber and for noise control. For applications like acoustic filters, sensors and sound lasers, acoustic resonances with ultrahigh quality-factor (Q-factor) are desired. Bound states in the continuum (BICs), also known as trapping mode with infinite large Q-factor, have triggered extensive interest in photonic and acoustic communities[5,6]. It corresponds to a non-radiating mode embedded within the continuum, enabling the enhanced light-matter interaction due to its excellent field confinement[7–9]. In fact, there is a long history of acoustic BICs. In 1951, Ursell derived the critical condition for the existence of BIC in an open acoustic system that consists of an entire sphere sandwiched between a circular waveguide[10]. Later, different types of BICs, including symmetry-protected BICs[11–14], Friedrich-Wintgen BICs[15–17], Fabry-Perot BICs[18], and mirror-induced BICs[19], have been reported theoretically with various acoustic systems. Also, some efforts have been made to demonstrate the existence of those BICs experimentally[19–23]. The highest reported Q-factor is up to 583[19]. However, most of the experimentally demonstrated BICs so far are quite sensitive to the perturbation of structure; that is, the Q-factor is inversely proportional to the perturbation square[24]. Thus, the following questions remain open: how to build topologically stable BICs (also called supercavity resonances or super-BICs) robust to fabrication imperfections[25].

This work presents a synergic effort of theoretical design and experimental verification of super-BICs in a coupled acoustic resonator system arising from merged BICs. Unlike topological BICs happening in the momentum space by periodic structures, the merged BIC effect is implemented in geometry parameter space (i.e., the width of the individual resonator and the gap between them) with a two coupled resonators system. It is attributed to topological charges merging and annihilation. Also, we unveil the subtle correlation between such supercavity resonances and Friedrich-Wintgen BICs in an entire rectangular (or cuboid)-waveguide system. We fabricated such kinds of coupled resonators and experimentally verified the merging effect of BICs by measuring their reflection spectra, which show excellent agreement with the numerical calculations. Notably, the effect of merged BICs is a universal phenomenon and can be extended to other systems, such as an electromagnetic wave. Our results offer promising applications in building high-performance acoustic and photonic devices, such as sensors, filters, and lasers.

## 2. Results and discussion

### 2.1 Merged BIC in 2D coupled acoustic resonators

We start by investigating the eigenmodes of the 2D coupled acoustic resonator, as schematically shown in Fig.1a-b. Our previous study has demonstrated that the leaky modes supported by such an open system play an essential role in governing the acoustic properties[19]. Each leaky mode can be represented by a complex eigenfrequency $\omega = \omega_0 - i\gamma$, where $\omega_0$ and $\gamma$ are respectively the resonant frequency and radiative decay rate. The radiative Q-factor can be obtained by $Q = \omega_0/2\gamma$. With the information of complex eigenfrequency for given modes, both reflection and transmission spectrum can be well reproduced based on temporal coupled-mode theory. Thus, the task of searching for BIC turns to find the leaky modes with zero radiative decay rate. For the sake of simplicity, we fix the waveguide and side resonator's height as $h_1 = 8cm$ and $h_2 = 20cm$, respectively. We first consider the case of coupled resonators with the width $L = 20cm$ while the distance between two resonators is varied. Such coupled resonators support bonding mode and antibonding mode, similar to the interaction of two molecules. Fig.1c-d show the pressure distribution of antibonding mode $M_{12a}$ and bonding mode $M_{12b}$. Previous studies indicated that these two modes become BIC (called as Fabry-Perot BIC[5,18]) only when the propagating phase between two resonators satisfies $\phi = k_0 d = m\pi$, where $k_0$ is the resonant wavenumber of single resonator and $d$ is the distance between two resonators. Based on this prediction, one should be able to find one anti-bonding and one bonding BICs for d≤ $\lambda_0$ ($\lambda_0$ is resonant wavelength of single resonator) and the distance difference should be close to $\lambda_0/2$. Quite surprisingly, we found from Fig.1f-g that there are two anti-bonding BICs and two bonding BICs. Importantly, the distance difference between two anti-bonding BICs (or bonding BICs) is less than $\lambda_0/2$ while the distance difference between first (second) anti-bonding BIC and first (second) bonding BIC is close to $\lambda_0/2$. The BICs will move toward each other if we further increase the resonators' width. When $L$ approaches the critical value $Lc = 20.302 cm$, a merged BIC is observed. For $L > Lc$, BIC vanishes and is reduced to quasi-BIC (QBIC). Furthermore, we found that for the structure that shows BIC before merging, the Q-factor of each BIC is inverse proportional to $1/(d - d_{BIC})^2$ (See Fig.S1a)[24]. When two antibonding (or bonding) BICs merge, the Q factor of each BIC is inverse proportional to $1/(d - d_{BIC})^4$ (See Fig.S1b). This phenomenon is similar to the case of merged BIC observed in the photonic system, where topological charges move toward the $\Gamma$ point in first Brillouin zone at momentum space[25–28]. The essential difference lies in that the merging effect is observed in momentum space for an array of photonic nanoresonators, but it happens in geometry parameter space for coupled acoustic resonators. The behavior of BICs moving, merging, and vanishing is confirmed by the reflection spectrum mapping of coupled resonators with $L = 20cm, 20.302cm, 20.50cm$ (See Fig.S2). This interesting phenomenon could be correlated to the strong near-field coupling of two resonators, evidenced by the rotation field distribution as shown in Fig.1c-d. To prove that strong near-field coupling indeed plays a vital role in the formation of super-BIC, we introduce a neck between the resonator and bus waveguide, as schematically drawn in Fig.1b. The Q-factor of bonding and antibonding modes vs distance are calculated and shown in Fig.1e and Fig.1h. Indeed, only one antibonding BIC and one bonding BIC, belonging to normal Fabry-Perot BICs, are observed as its Q-factor is inverse proportional to $1/(d - d_{BIC})^2$ (See Fig.S3a). This phenomenon is also confirmed by the reflection mapping shown in Fig.S3b.

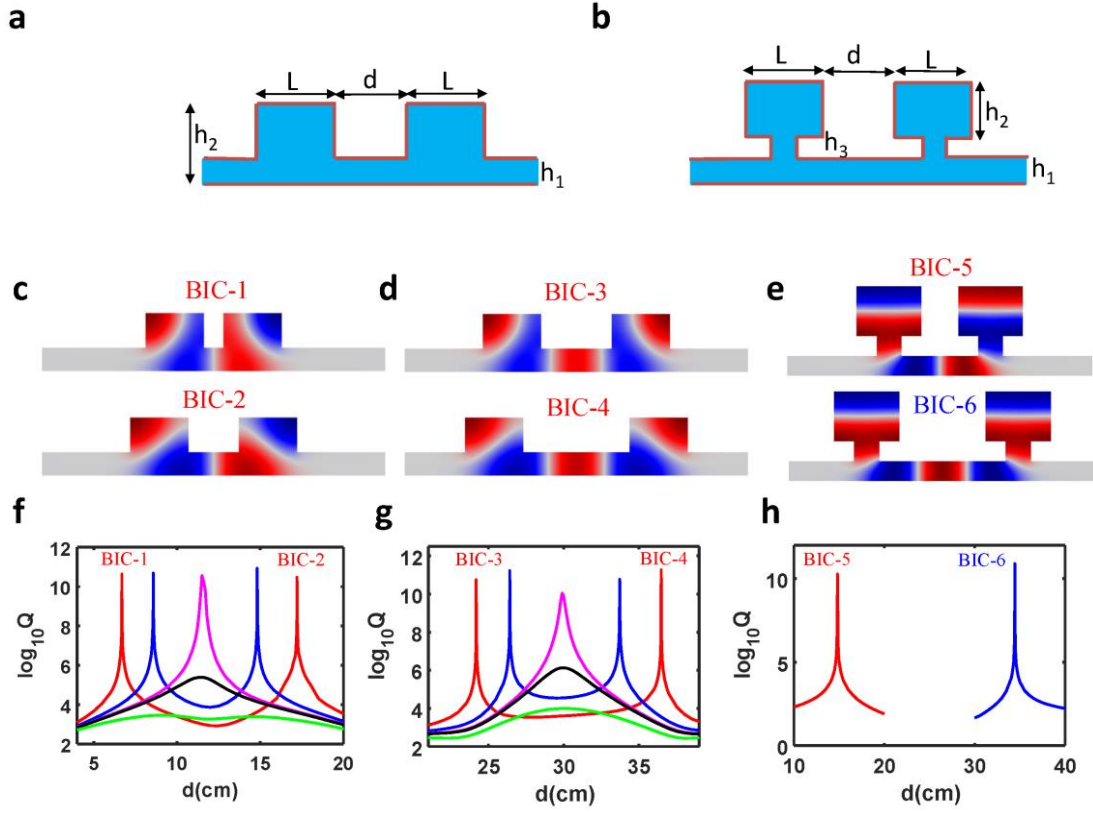

**Figure 1. BICs in the coupled-resonator system**. a, schematic drawing of the coupled-resonator without neck. b, schematic drawing of the coupled resonators with a neck. c, Pressure field distribution for two antibonding BICs. d, Pressure field distribution for two bonding BICs. e, Pressure field distribution for antibonding and bonding BICs for system with neck. f, Q-factor for antibonding mode as a function of distance between two resonators for width L=20cm (red), L=20.2cm (blue), L=20.302cm (magenta), L=20.32cm(black), L=20.5cm(green) while the heigh of single resonator and waveguide is fixed as 20cm and 8cm, respectively. g, Q-factor for bonding mode as a function of distance between two resonators for width L=20cm (red), L=20.2cm (blue), L=20.302cm (magenta), L=20.32cm(black), L=20.5cm(green). h, Q-factor for both bonding and antibonding modes as a function of the distance between two resonators. The width and height of each resonator are L=26cm and $h_2$=20cm. The height of the waveguide is 8cm. The neck width is 10cm and height is 8cm.

Importantly, such a merging effect can be well explained by the temporal coupled-mode theory. Fig.2a shows the schematic drawing of such a coupled-resonators system, whose Hamiltonian takes the following general forms

$$H = \begin{pmatrix} \omega_0 - i\gamma & C_1 + iC_2 \\ C_1 + iC_2 & \omega_0 - i\gamma \end{pmatrix} \quad (1)$$

where $\omega_0$ and $\gamma$ are the resonant frequency and radiative decay rate, respectively. $C_1$ and $C_2$ are the real and imaginary parts of complex coupling between two resonators. Solving Eq. (1) gives us

$$\omega_+ = \omega_0 + C_1 - i(\gamma - C_2) \quad (2a)$$

$$\omega_- = \omega_0 - C_1 - i(\gamma + C_2) \qquad (2b)$$

Where $\omega_+$ and $\omega_-$ are, respectively, the complex eigenfrequencies of bonding and antibonding modes. From Eq. (2), we can find that bonding (antibonding) modes become BIC for $C_2 = \pm\gamma$, while $C_1$ plays a minor role in the BICs' formation. Previous studies have demonstrated that such a coupled-resonator system support Fabry-Perot BICs[5] by assuming $C_1 = \kappa + \gamma\sin(k_0 d)$ and $C_2 = -\gamma\cos(k_0 d)$, where $k_0 = \omega_0/c$ and $\kappa$ is the near-field coupling term. When the phase accumulation $k_0 d = m\pi$, either bonding or antibonding mode reduces to an ideal BIC.

However, they are not the only complex coupling terms that produce the Fabry-Perot BICs. For the sake of simplicity, we assume bonding and antibonding BICs occur at $k_0 d_1 = \pi$ and $k_0 d_2 = 2\pi$. Except for $C_2 = -\gamma\cos(k_0 d)$, we found that FP-BICs could be successfully constructed by setting $C_2 = \gamma[-1 + A(k_0 d - \pi)^2]$ ($\pi/2 < k_0 d \leq 3\pi/2$) and $C_2 = \gamma[1 - A(k_0 d - 2\pi)^2]$ ($3\pi/2 < k_0 d \leq 5\pi/2$), where $A = 1/(1.5\pi - \pi)^2$.

Fig.2b shows $C_2$ as a function of the propagation phase $k_0 d$. We also plot the function $C_2 = -\gamma\cos(k_0 d)$ to make a comparison. Surprisingly, we find that these two functions almost overlap with each other. Additionally, since the real part of the coupling term shows no contribution to the formation of BICs, we assume $C_1 = 10Hz$. Also, without losing generality, $\omega_0$ and $\gamma$ are fixed as 1000Hz and 10Hz, respectively. Substituting these parameters into Eq. (2) gives us the complex eigenfrequencies of bonding and antibonding modes, which are plotted in Fig.2c. Due to the similar $C_2$ functions, bonding (antibonding) modes have almost the same Q-factors. Note that around both BICs their Q-factors are indeed inversely proportional to $(k_0\Delta d)^2$.

After having this knowledge, it is straightforward to construct the complex coupling term for merged BICs cases. We still assume $C_1$=10Hz, $\omega_0$=1000Hz, and $\gamma$=10Hz, while $C_2$ is written as follows

$$C_2(k_0 d) = \gamma[-A + B(k_0 d - k_0 d_1)^2(k_0 d - k_0 d_2)^2] \quad (0 < k_0 d \leq \pi) \quad (3a)$$

$$C_2(k_0 d) = \gamma[A - B(k_0 d - k_0 d_3)^2(k_0 d - k_0 d_4)^2] \quad (\pi < k_0 d \leq 2\pi) \quad (3b)$$

where A=1 before BIC merging and BIC merging, and A=1-δ after BIC merging. Here, δ is a small perturbation term that converts BIC into QBIC after BIC merging. B can be obtained based on the continuity condition at $k_0 d = \pi$. It is necessary to point out that there is an additional requirement $k_0 d_3 = k_0 d_1 + \pi$ and $k_0 d_4 = k_0 d_2 + \pi$.

Fig.2d shows an example of $C_2$ built from Eq. (3) with different $k_0d_1$ by assuming $k_0d_1 + k_0d_2 = \pi$. Fig.2e-f shows the Q-factors of both antibonding and bonding modes as a function of distance. It is evident that this simple model has successfully predicted BIC moving, merging, and vanishing. Also, the Q-factors for both modes around BICs are inversely proportional to $(k_0\Delta d)^2$ before BIC merging. When $k_0d_1 = k_0d_2 = \pi/2$ and $k_0d_3 = k_0d_4 = 3\pi/2$, two antibonding (bonding) BICs merge and their Q-factors are now inversely proportional to $(k_0\Delta d)^4$. Note that this $C_2$ function is not the only one that could generate merged BICs. Another effective $C_2$ function could be easily constructed by replacing $(k_0d - k_0d_i)^2$ with $[1 - \cos(k_0d - k_0d_i)](i = 1,2,3,4)$ (See Fig.S4).

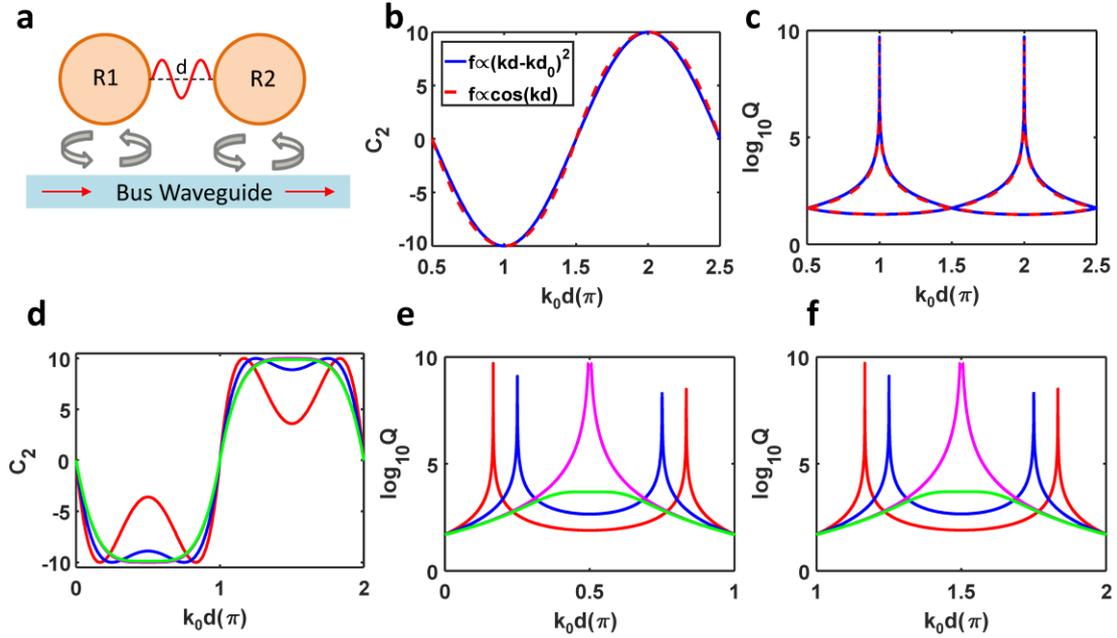

**Figure 2. Coupled mode theory for merged BICs**. a, schematic drawing of the coupled-resonators system via a bus waveguide. b, the imaginary part of coupling $C_2$ vs propagation phase $k_0d$ for normal Fabry-Perot BICs. c, Q-factors of bonding and antibonding modes as a function of propagation phase $k_0d$. d, the imaginary part of coupling $C_2$ vs propagation phase $k_0d$ for merged BICs cases. Note that red line represents $k_0d_1=\pi/6$ and $k_0d_2=5\pi/6$, blue line represents $k_0d_1=\pi/4$ and $k_0d_2=3\pi/4$, magenta line represents $k_0d_1= k_0d_2=\pi/2$, green line represents $k_0d_1= k_0d_2=\pi/2$ and A=0.99. e-f, Q-factors of antibonding mode (e) and bonding mode (f) vs $k_0d$.

## 2.2 General merged BICs in 2D coupled acoustic resonators.

Note that the case in Fig.1f-g is not the only structure that shows BIC merging effect. One could construct many merged BICs by checking the Friedrich-Wintgen BIC in an open system[17,19,29]. Here, we consider two open acoustic systems, as shown in Fig.3a. One is made of an acoustic cavity connecting to two waveguides with left and right open boundaries. The other corresponds to an acoustic cavity coupled to a single port waveguide. As demonstrated in our previous work, some of the BICs supported in these two systems share similar features due to the mirror-effect[19]. Taking mode $M_{13}$ as an example, from Fig.3b-c, we indeed found that both structures support a BIC at critical size ratio $Rc \approx 0.992$ while the pressure distribution of BICs is one-quarter of the pressure distribution for the former one owning to the hard wall boundary induced mirror effect. Moreover, the eigenfield profile of BIC for the single cavity-single port waveguide system shares similar field distribution of half in coupled resonators. This suggests that one may find more merged BICs by constructing a Friedrich-Wintgen BIC with eigenfield profile that possesses symmetry along both x- and y-axes in the full cavity-waveguide system. We confirm this hypothesis by finding more examples on merged BIC in such a coupled resonators system. Here, we show one of them in Fig.3d-f, where the height of waveguide is $h_1 = 5cm$ and the height of resonator is $h_2 = 30cm$. The red, blue, magenta, and black lines represent the case of coupled resonators with the width $L = 20cm, 20.4cm, 20.674cm\ and\ 20.7cm$, respectively. Note that only antibonding BICs are shown here to illustrate the merging effect and results of bonding BICs are given in supporting information (See Fig.S5). From Fig.3f, antibonding BICs indeed experience moving, merging, and vanishing with the increasing width of rectangular cavity but a fixed height. However, the left and right BICs have different moving speed toward the merged point due to the asymmetric distribution with respect to the diagonal of single resonator. More examples can be found in supporting information (Figs.S6-7). Besides, by taking advantage of mirror effect, we found that even single resonator with mirror on the right boundary could support such a merging BICs, which has been demonstrated in Fig.3g-i. In addition, as pointed out in previous section, the common feature of such BICs is that they have rotated pressure distribution with respect to x axis, which indicates a strong near-field coupling. Thus, we can construct merged BIC in different coupled resonator system that also ensures the strong near-field coupling. The corresponding results are summarized in Fig.S8.

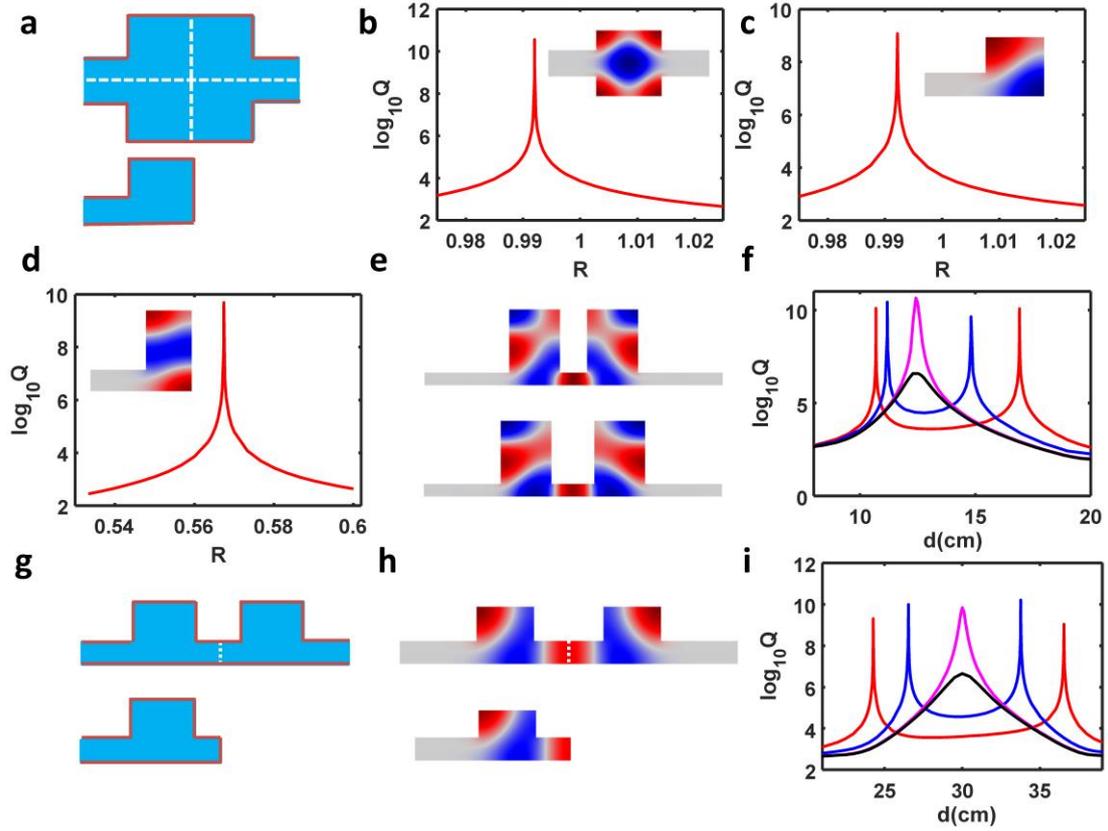

**Figure 3. General features of merged BICs**. a, schematic drawing of a full-rectangular resonator sandwiched between two rectangular waveguides (top panel) and a quarter-rectangular resonator connected to single-port rectangular waveguide (bottom panel). b, Q-factor vs size ratio R (R=Ly/Lx) for the full-resonator system. The inset shows the pressure field distribution of Friedrich-Wintgen BIC. c, Q-factor vs size ratio R (R=Ly/Lx) for the quarter-resonator system. The inset shows the pressure field distribution of BIC. d, another example of BIC supported by a quarter-resonator-single-port waveguide system. Q-factor vs size ratio R (R=Ly/Lx) for the quarter-resonator system. The inset shows the pressure field distribution of this BIC. e, Pressure field distribution for two antibonding BICs. f, Q-factor for antibonding mode as a function of distance between two resonators for width L=20cm (red), 20.4cm(blue), 20.674cm(magenta) and 20.7cm(black). g, Schematic drawing of coupled-resonator and single resonator with mirror. h, pressure field distribution of bonding BICs in a coupled resonators system (up) and BIC in single resonator-single port system (bottom). i, Q-factor for bonding mode as a function of distance between resonator and right boundary of system for width L=20cm(red), L=20.2cm(blue), L=20.301cm(magenta), L=20.31cm(black).

## 2.3 Topological origin of the merged BICs

The merging effect of two BICs in such a coupled-resonator system could be perfectly explained from a topological perspective. Unlike the case of photonic BIC, where each BIC carries a topological charge in momentum space, we found that each BIC in such a finite acoustic system is represented by a pair of topological charges with one $q = +1$ and the other $q = -1$[30]. Without loss of generality, we consider the single-resonator-single port system shown in Fig.3g as an example and set the resonator's height and width as 20cm. From Fig.3i, it can be seen that there are two BICs when the distance changes from 20cm to 40cm. To define the topological charge of BICs, an ultrasmall artificial loss has been introduced in the sound velocity to quench the reflection amplitude to zero (or perfect absorption) by critical coupling. As the reflection is reduced to zero, its phase could not be clearly defined and thus could be viewed as a phase-singularity. Here, it is worth noting that there are always two phase-singularities around each BIC when a slight loss is intentionally introduced. The topological charge that each phase-singularity carries is defined as the accumulated phase when a closed path encircles the phase-singularity counter-clockwise in the distance-frequency plane:

$$q = \frac{1}{2\pi} \oint d\varphi \tag{4}$$

Indeed, there are two zero-reflection points in amplitude mapping, as shown in Fig.4a-b. Each zero-reflection point corresponds to a phase vortex in the phase mapping of Fig.4e-f. Then, we calculate the phase along a closed path encircling the phase-vortex indicated by a yellow rectangle in Fig.4a-b and plot it in Fig.S9. By using Eq. (4), the topological charges can be evaluated to $q_1 = +1$ and $q_2 = -1$ for the two-phase vortices, respectively. Similarly, the second BIC carries topological charge pair $q_3 = +1$ and $q_4 = -1$. Thus, we can conclude that each BIC could be correlated to a pair of opposite topological charges. When the width of the resonator increases to 20.301cm, these two BICs merge and are reduced to a single BIC, as confirmed by Fig.4c and Fig.4g. Also, topological charge analysis indicates that the middle two topological charges coalesce and exhibit and only one pair of topological charges is left (See Fig.S10), indicating the merging BICs. Further increasing the resonator's width leads to the vanishing of BIC. No topological charge will be found for this structure because the critical coupling is no longer satisfied, as verified in Fig. 4d and Fig.4h. Fig. 4i summarizes the entire process of BIC moving, merging, and vanishing by computing the topological charge evolution. Two pairs of topological charges move to each other, annihilate, and disappear finally.

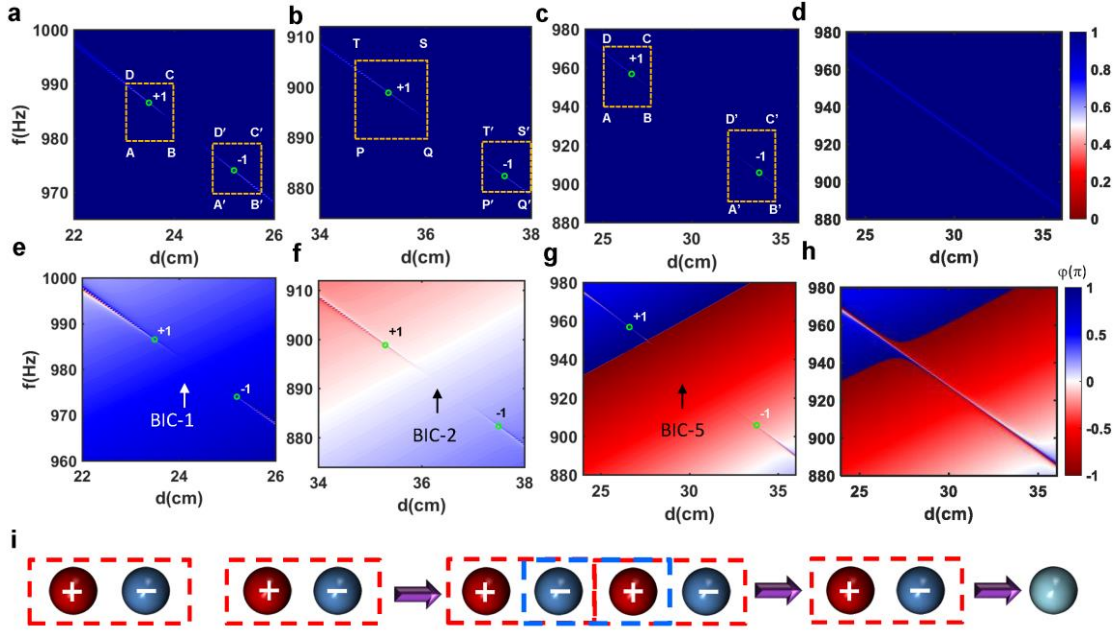

**Figure 4. Topological explanation of merged BICs**. a-b, Reflection mapping of the single resonator with a mirror around BIC-1 (a) and BIC-2 (b) before merging. An artificial loss is introduced in the sound velocity (v=343*(1+2.5e-5i)). c, reflection mapping for merged BIC. d, reflection mapping after BIC merging. e-f, phase mapping of the single resonator with a mirror around BIC-1(e) and BIC-2 (f). g, phase mapping for merged BIC. h, phase mapping after BIC merging. i, topological charge evolution process before and after BIC merging.

## 2.4 Experimental demonstration of the merged BIC in 3D coupled-resonators.

Such merged BICs also happen in 3D cases, where two cuboid resonators coupled with each other via an acoustic waveguide (e.g., a circular waveguide or a rectangular waveguide), as schematically shown in Fig.5a. Here, the diameter of the waveguide is 29mm for the sake of experimental consideration. The length and height of the cuboid resonator are fixed as 29mm and 60mm, respectively. By varying the length of the cuboid resonator from L=60mm, 61mm, 62,212mm to 63mm, from Fig.5a-c, we indeed observe the similar effects of BICs moving, merging, and vanishing when the distance between coupled resonators is continuously changed. Similar to 2D cases, we can find more merged BICs by constructing Friedrich-Wintgen BICs, as shown in Fig.S11-13. To experimentally verify the merging BIC, we fabricate a series of coupled resonators-waveguide structures. Fig.5d shows the optical images of two printed samples. The lengths of each resonator are chosen as L=61mm, 62.2mm, 63mm, while the distance between resonators covers a broad range from 10mm to 140mm. Fig.5e-g shows the measured reflection mapping, while Fig.5h-j displays the calculated reflection spectrum. Excellent agreement is found between the experiments and simulations. For L=61mm, there are two antibonding BICs and two bonding BICs. When L is increased to 62.2mm, both bonding BICs and antibonding BICs start merging. Further increasing L leads to the vanishing of BIC.

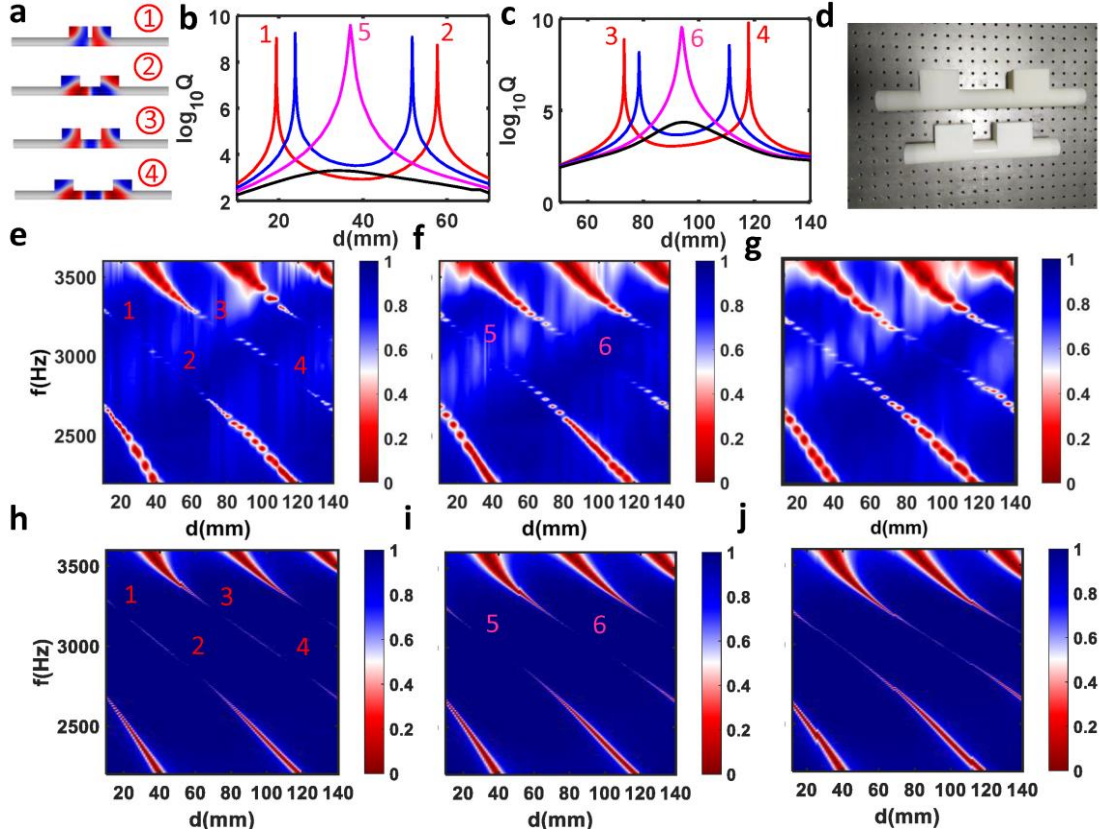

**Figure 5. Experimental demonstration of merged BICs in 3D coupled resonator system**. a, Pressure field distribution for two antibonding BICs (#1-2)and two antibonding BICs (#3-4). b, Q-factor for antibonding mode as a function of distance between two resonators for width L=61mm(red), 61.6mm(blue), 62.212mm(magenta), 63mm(black). The diameter of circular waveguide is 29mm.c, Q-factor for bonding mode as a function of distance between two resonators for width L=61mm(red), 61.6mm(blue), 62.212mm(magenta), 63mm(black). d. Optical image of 3D coupled resonator system fabricated by 3D-printing. (e-g). Measured reflection spectrum mapping vs distance between two resonators for the resonator's width L=61mm (e), L=62.2mm (f), and L=63mm(g). (h-j). Simulated reflection spectrum mapping vs distance between two resonators for the resonator's width L=61mm(h), L=62.2mm(i), and L=63mm(j).

## 2.5 Generalization of merged BIC to photonic waveguide system

Finally, it is worth commenting that such merged BICs are not limited to the acoustic wave. It could also be generalized to other systems supporting wave-like excitations, such as electromagnetic structures. To construct merged BICs, one just needs to replace the rigid wall boundary of a coupled acoustic resonator system with either perfect electric conductor (PEC) or perfect magnetic conductor (PMC), depending on the polarization choice. The examples are shown in Fig.S14 for transverse electric (TE) polarization and Fig.S15 for transverse magnetic (TM) polarization in the supplementary materials. Thus, we can envision many exciting applications that could be developed based on such a coupled photonic resonator system, such as a high-performance filter at the microwave range.

## 3. Conclusion

We proposed the theoretical design and experimental demonstration of supercavity resonances in two coupled acoustic resonators arising from merged BICs. Their formation could be perfectly explained by their topological charges' evolution. Additionally, we found that strong near-field coupling plays an essential role in constructing such merged BICs. We also unveiled the general rules of finding out such kinds of supercavity resonances, which could be rigorously correlated to the Friedrich-Wintgen BIC in an entire rectangular resonator with two ports connecting to waveguide. Following this rule, we could construct two equivalent systems which are also able to support supercavity resonances. Moreover, we demonstrate that the topological characteristics of the merging effect are explained by the topological charges carried by the reflection phase singularity. Finally, we fabricated a series of samples and demonstrated their peculiar property in two coupled 3D resonators-BICs moving, merging, and vanishing. Our results may find applications in constructing high-performance acoustic devices, such as acoustic filters, acoustic sensors, acoustic lasers, etc.

## Materials and Methods

### Simulations

Both the eigenmodes and reflection/transmission spectra are performed with the commercial software COMSOL Multiphysics. The speed of sound and air density is set as 343 m/s and 1.29kg/m$^3$, respectively. When calculating eigenmodes and reflection spectrum, perfect matched layer boundaries at the two ends of waveguides are applied to mimic acoustic wave propagation in the infinite space. The other exterior boundaries are set as hard walls.

## Experiments

We fabricate these coupled acoustic resonators with 3D-printing technology using laser sintering stereolithography (SLA, 140μm) with a photosensitive resin (UV curable resin). Reflection spectra of these samples are measured using a Brüel & Kjær type-4206T impedance tube with a diameter of 29 mm. A loudspeaker generates a plane wave, and the amplitude and phase of local pressure are measured by four 1/4-inch condenser microphones (Brüel & Kjær type-4187) situated at designated positions.

## Acknowledgements

L. Huang and A. E. Miroshnichenko were supported by the Australian Research Council Discovery Project (DP200101353) and the UNSW Scientia Fellowship program. Y. K. Chiang and D. A. Powell were supported by the Australian Research Council Discovery Project (DP200101708), B. Jia, S. Huang, and Y. Li were supported by the National Natural Science Foundation of China (Grants No. 12074286).

## Author Contributions

L. H and A. E. M conceived the idea. L. H performed the theoretical calculation and numerical simulation. Y. K. C, and C. S, F. D and T. Y help with numerical simulations. B. J, S. H and Y. L fabricated the sample and performed the reflection/transmission spectra measurements. L. H, Y. L, and A. E. M supervised the project. L. H and A. E. M prepared the manuscript with the input from all authors.